\documentstyle[12pt]{article}
\input psfig.sty
\font\tenrm=cmr10
\font\tenit=cmti10
\font\elevenbf=cmbx10 scaled\magstep 1
 1
 1

\def\nn{\noindent}
\def\Re{{\cal R \mskip-4mu \lower.1ex \hbox{\it e}\,}}
\def\Im{{\cal I \mskip-5mu \lower.1ex \hbox{\it m}\,}}

\def\etal{{\it et al.}}

\def\sub#1{_{\lower.25ex\hbox{$\scriptstyle#1$}}}
\def\sul#1{_{\kern-.1em#1}}
\def\sll#1{_{\kern-.2em#1}}
\def\sbl#1{_{\kern-.1em\lower.25ex\hbox{$\scriptstyle#1$}}}
\def\ssb#1{_{\lower.25ex\hbox{$\scriptscriptstyle#1$}}}
\def\sbb#1{_{\lower.4ex\hbox{$\scriptstyle#1$}}}

\def\gev{\,{\rm GeV}}

\def\to{\rightarrow}
\def\mh{\ifmmode m\sbl H \else $m\sbl H$\fi}
\def\mch{\ifmmode m_{H^\pm} \else $m_{H^\pm}$\fi}
\def\mt{\ifmmode m_t\else $m_t$\fi}
\def\mc{\ifmmode m_c\else $m_c$\fi}
\def\mz{\ifmmode M_Z\else $M_Z$\fi}
\def\mw{\ifmmode M_W\else $M_W$\fi}
\def\mws{\ifmmode M_W^2 \else $M_W^2$\fi}
\def\mhs{\ifmmode m_H^2 \else $m_H^2$\fi}
\def\mzs{\ifmmode M_Z^2 \else $M_Z^2$\fi}
\def\mts{\ifmmode m_t^2 \else $m_t^2$\fi}
\def\mcs{\ifmmode m_c^2 \else $m_c^2$\fi}
\def\mchs{\ifmmode m_{H^\pm}^2 \else $m_{H^\pm}^2$\fi}
\def\ztwo{\ifmmode Z_2\else $Z_2$\fi}
\def\zone{\ifmmode Z_1\else $Z_1$\fi}
\def\mtwo{\ifmmode M_2\else $M_2$\fi}
\def\mone{\ifmmode M_1\else $M_1$\fi}
\def\tb{\ifmmode \tan\beta \else $\tan\beta$\fi}
\def\xw{\ifmmode x\sub w\else $x\sub w$\fi}
\def\ch{\ifmmode H^\pm \else $H^\pm$\fi}
\def\lum{\ifmmode {\cal L}\else ${\cal L}$\fi}
\def\inpb{\ifmmode {\rm pb}^{-1}\else ${\rm pb}^{-1}$\fi}
\def\infb{\ifmmode {\rm fb}^{-1}\else ${\rm fb}^{-1}$\fi}
\def\epem{\ifmmode e^+e^-\else $e^+e^-$\fi}
\def\ppb{\ifmmode \bar pp\else $\bar pp$\fi}
\def\bkg{\ifmmode B\to K^*\gamma\else $B\to K^*\gamma$\fi}
\def\bsg{\ifmmode b\to s\gamma\else $b\to s\gamma$\fi}

\newskip\zatskip \zatskip=0pt plus0pt minus0pt
\def\matth{\mathsurround=0pt}

\def\atversim#1#2{\lower0.7ex\vbox{\baselineskip\zatskip\lineskip\zatskip
  \lineskiplimit 0pt\ialign{$\matth#1\hfil##\hfil$\crcr#2\crcr\sim\crcr}}}

\renewenvironment{thebibliography}[1]
 { \tenrm
   \begin{list}{\arabic{enumi}.}
    {\usecounter{enumi} \setlength{\parsep}{0pt}
     \setlength{\itemsep}{3pt} \settowidth{\labelwidth}{#1.}
     \sloppy
    }}{\end{list}}

\begin{document}
\rightline{\vbox{\halign{&#\hfil\cr
&SLAC-PUB-95-6782\cr
&February 1995\cr}}}
\begin{center}{{\elevenbf PROBING NEW PHYSICS IN B PENGUINS}
\footnote{Work Supported by the Department of Energy,
Contract DE-AC03-76SF00515}
\footnote{Presented at the {\it International Workshop on B Physics: Physics
Beyond the Standard Model at the B Factory}, Nagoya, Japan, October 26-28,
1994}
\\
\vglue 0.6cm
{\tenrm J.L. HEWETT \\}
\baselineskip=13pt
{\tenit Stanford Linear Accelerator Center \\}
\baselineskip=12pt
{\tenit Stanford University, Stanford, CA   94309\\}
\vglue 0.2cm
{\tenrm ABSTRACT}}
\end{center}
\vglue 0.05cm
{\rightskip=3pc
 \leftskip=3pc
 \tenrm\baselineskip=12pt
 \noindent
Constraints placed on physics beyond the Standard Model from the recent CLEO
observation of the inclusive decay $B\to X_s\gamma$ are summarized.  Further
searches for new physics using the process $B\to X_s\ell^+\ell^-$ are
discussed.
\vglue 0.6cm}
The study of virtual effects can open an important window on electroweak
symmetry breaking and physics beyond the Standard Model (SM).  The examination
of indirect effects of new physics in higher order processes offers a
complementary approach to the search for direct production of new particles
at high energy colliders.  In fact, tests of loop induced couplings can
provide a means of probing the detailed structure of the SM at the level
of radiative corrections where Glashow-Iliopoulos-Maiani (GIM) cancellations
are important.  In some cases the constraints on new degrees of freedom
via indirect effects surpass those obtainable from collider searches.  In
other cases, entire classes of models are found to be incompatible.  Given
the amount of high luminosity data on the $B$ system which will become
available during the next decade, this approach to searching for physics
beyond the SM will become an increasingly valuable tool.

Radiative $B$ decays have become one of the best testing grounds of the SM
due to recent progress on both the experimental and theoretical fronts.  The
CLEO Collaboration has recently reported\cite{cleoin} the observation of
the inclusive decay $B\to X_s\gamma$ with a branching fraction of
$(2.32\pm 0.57\pm 0.35)\times 10^{-4}$.  Observation of this process at the
inclusive level removes the uncertainties associated with folding in the
imprecisely predicted\cite{soni} ratio of exclusive to inclusive rates
when comparing theoretical results with exclusive data.  On the theoretical
side, the reliability of the calculation of the quark-level process \bsg\
is improving\cite{qcd} as agreement on the leading-logarithmic QCD corrections
has been reached and partial calculations at the next-to-leading logarithmic
order are underway.  These new results have inspired a large number of
investigations of this decay in various classes of models, which can be
summarized by the following list:
\pagebreak

\begin{itemize}
\item ``Top Ten'' Models Probed by \bsg
\begin{tabbing}
3. Anomalous Trilinear Gauge Couplings extended \= 8. Extended Technicolor
\kill
1. Standard Model  \> 6. Supersymmetry\\
2. Anomalous Top-Quark Couplings  \> 7. Three-Higgs-Doublet Model\\
3. Anomalous Trilinear Gauge Couplings \>  8. Extended Technicolor\\
4. Fourth Generation \>  9. Leptoquarks\\
5. Two-Higgs-Doublet Models \>  10. Left-Right Symmetric Models
\end{tabbing}
\end{itemize}
Clearly, I only have time to discuss a couple of these models here,
a more complete review can be found in Ref. \cite{ssi}.

In the SM, the quark-level transition \bsg\ is mediated by $W$-boson and
t-quark exchange in an electromagnetic penguin diagram.  To obtain
the branching fraction, the inclusive rate is scaled to that of the
semi-leptonic decay $b\to X\ell\nu$.  This procedure removes uncertainties
in the calculation due to an overall factor of $m_b^5$ which appears in
both expressions, and reduces the ambiguities involved with the imprecisely
determined Cabibbo-Kobayashi-Maskawa (CKM) factors.  The result is then
rescaled by the experimental value\cite{pdg} of
$B(b\to X\ell\nu)$.  The semi-leptonic rate is calculated incorporating
both charm and non-charm modes, and includes both phase space and QCD
corrections\cite{nicolo}.  The calculation of $\Gamma(\bsg)$ employs the
renormalization group evolution\cite{qcd} for the
coefficients of the $b\to s$ transition operators in the effective Hamiltonian
at the leading logarithmic level.  The participating operators consist of
the current-current operators $O_{1,2}$, the QCD penguin operators $O_{3-6}$,
and the electro- and chromo-magnetic operators $O_{7,8}$.
The Wilson coefficients of the $b\to s$ operators are evaluated perturbatively
at the $W$ scale, where the matching conditions are imposed, and evolved
down to the renormalization scale $\mu$, usually taken to be $\sim m_b$.
This
procedure yields the branching fraction $B(\bsg)=2.92^{+0.77}_{-0.59}\times
10^{-4}$ for a top-quark mass of 175 GeV.  The central value corresponds to
$\mu=m_b$, while the upper and lower errors represent the deviation due to
assuming $\mu=m_b/2$ and $\mu=2m_b$, respectively.  We see that (i) this value
compares favorably to the recent CLEO measurement and (ii) the freedom of
choice in the value of the renormalization scale introduces an uncertainty
of order $25-30\%$.  Clearly, when determining
constraints on new physics from this process, one must choose values for
the parameters which yield the most conservative limits.

Before discussing explicit models of new physics, we first investigate the
constraints placed directly on the Wilson coefficients of the magnetic moment
operators.  Writing the coefficients at the matching scale in the form
$c_i(M_W)=c_i(M_W)_{SM}+c_i(M_W)_{new}$, where $c_i(M_W)_{new}$ represents
the contributions from new interactions, we see that the CLEO measurement
limits the possible values of $c_i(M_W)_{new}$ for $i=7,8$.  These bounds
are presented in Fig. 1(a), for $m_t=175\gev$,
where the allowed regions lie inside the diagonal
bands.  We note that the two bands occur due to the overall sign ambiguity in
the determination of the coefficients (recall that $B(\bsg)\propto
|c_7^{eff}(\mu)|^2$), and by including the upper and lower CLEO bounds.  The
horizontal lines correspond to potential limits $B(b\to sg)<(3-30)\times
B(b\to sg)_{SM}$.
We see that such a constraint on $b\to sg$ is needed to further restrict the
values of the Wilson coefficients.

The trilinear gauge coupling of the photon to $W^+W^-$ can be tested
in \bsg\ decay.  Anomalous $\gamma WW$ vertices can be probed by
looking for deviations from the SM in tree-level processes such as
$\epem\to W^+W^-$ and $p\bar p \to W\gamma$, or by their influence on
loop order processes, for example the $g-2$ of the muon.  In the latter case,
cutoffs must be used in order to regulate the divergent loop integrals
and can introduce errors by attributing a physical significance to the
cutoff\cite{burgess}.  However, some loop processes, such as \bsg,
avoid this problem due to cancellations provided by the GIM mechanism,
and hence yield cutoff independent bounds on anomalous couplings.
The CP-conserving interaction Lagrangian for $WW\gamma$ interactions is
\begin{equation}
{\cal L}_{WW\gamma}= i\left( W^\dagger_{\mu\nu}W^\mu A^\nu-W^\dagger_\mu
A_\nu W^{\mu\nu}\right) +i\kappa_\gamma W^\dagger_\mu W_\nu A^{\mu\nu}
+i{\lambda_\gamma\over\mws}W^\dagger_{\lambda\mu}W^\mu_\nu A^{\nu\lambda}\,,
\end{equation}
where $V_{\mu\nu}=\partial_\mu V_\nu-\partial_\nu V_\mu$, and the two
parameters $\kappa_\gamma=1+\Delta\kappa_\gamma$ and $\lambda_\gamma$ take on
the values $\Delta\kappa_\gamma, \lambda_\gamma=0$ in the SM.  In this case,
only the coefficient
of the magnetic dipole operator, $O_7$, is modified
by the presence of these additional terms and can be written as
\begin{equation}
c_7(M_W) = G_7^{SM}(\mts/\mws) +\Delta\kappa_\gamma A_1(\mts/\mws)
+\lambda_\gamma A_2(\mts/\mws) \,.
\end{equation}
The explicit form of the functions $A_{1,2}$ can be found in
Ref. \cite{tgr}.  As both of these parameters are varied, either large
enhancements or suppressions over the SM prediction for the \bsg\ branching
fraction can be obtained.  When one demands consistency with both the upper
and lower CLEO bounds, a large region of the
$\Delta\kappa_\gamma-\lambda_\gamma$ parameter plane is excluded; this
is displayed in Fig. 1(b) from Ref. \cite{cleoin} for $m_t=174\gev$.
Here, the allowed region is given by the cross-hatched area, where
the white strip down the middle is excluded by the
lower bound and the outer white areas are ruled out by the upper limit on
$B(\bsg)$.  The ellipse represents the region allowed by
D0\cite{dzero}.
Note that the SM point in the $\Delta\kappa_\gamma-
\lambda_\gamma$ plane (labeled by the dot) lies in the center of one of the
allowed regions.
We see that the collider constraints are complementary to those from \bsg.

Next we turn to two-Higgs-doublet models (2HDM), where we examine the case
(denoted as Model II)
where the second doublet, $\phi_2$, gives mass to the up-type quarks, while the
down-type quarks and charged leptons receive their mass from $\phi_1$.  Each
doublet obtains a vacuum expectation value (vev) $v_i$, subject to the
constraint that $v_1^2+v_2^2=v^2$, where $v$ is the usual vev present in the
SM.  The charged Higgs boson interactions with
the quark sector are governed by the Lagrangian
\begin{equation}
{\cal L}= {g\over 2\sqrt 2M_W} \ch\left[ V_{ij}m_{u_i}A_u\bar u_i(1-\gamma_5)
d_j+V_{ij}m_{d_j}A_d\bar u_i(1+\gamma_5)d_j \right] + H.c. \,,
\end{equation}
where $g$ is the usual SU(2) coupling constant and $V_{ij}$ represents the
appropriate CKM element.  In Model II, $A_u=\cot\beta$ and $A_d=\tan\beta$,
where $\tan\beta\equiv v_2/v_1$ is the ratio of vevs.   The
\ch\ contributes to \bsg\ via virtual
exchange together with the top-quark and the dipole $b\to s$ operators
($O_{7,8}$) receive contributions from this exchange.  At the $W$ scale the
coefficients of these operators take the form
\begin{equation}
c_i(M_W)=G_i^{SM}(\mts/\mws)+A_{1_i}^{\ch}(\mts/\mchs)
+{1\over \tan^2\beta}A_{2_i}^{\ch}(\mts/\mchs) \,,
\end{equation}
where $i=7,8$.
The analytic
form of the functions $A_{1_i}, A_{2_i}$ can be found in \cite{bsgch}.
In Model II, large enhancements appear for small values of \tb, but
more importantly, we see that $B(\bsg)$ is always larger than that of the SM,
independent of the value of \tb\ due to the presence of
the $A_{1_i}^{\ch}$ term.  In this case, the CLEO upper bound
excludes\cite{cleoin,me}
the region to the left and beneath the curves shown in Fig. 1(c) for
$m_t=174\pm 16\gev$ and $\mu=2m_b$.  We note that the \ch\ couplings present in
Model II are of the type present in Supersymmetry.   However, the limits
obtained in supersymmetric theories also depend on the size of the other
super-particle contributions to \bsg, and are generally much more
complex\cite{bert,okada}.

\vspace*{-0.5cm}
\nn
\begin{figure}[htbp]
\centerline{
\psfig{figure=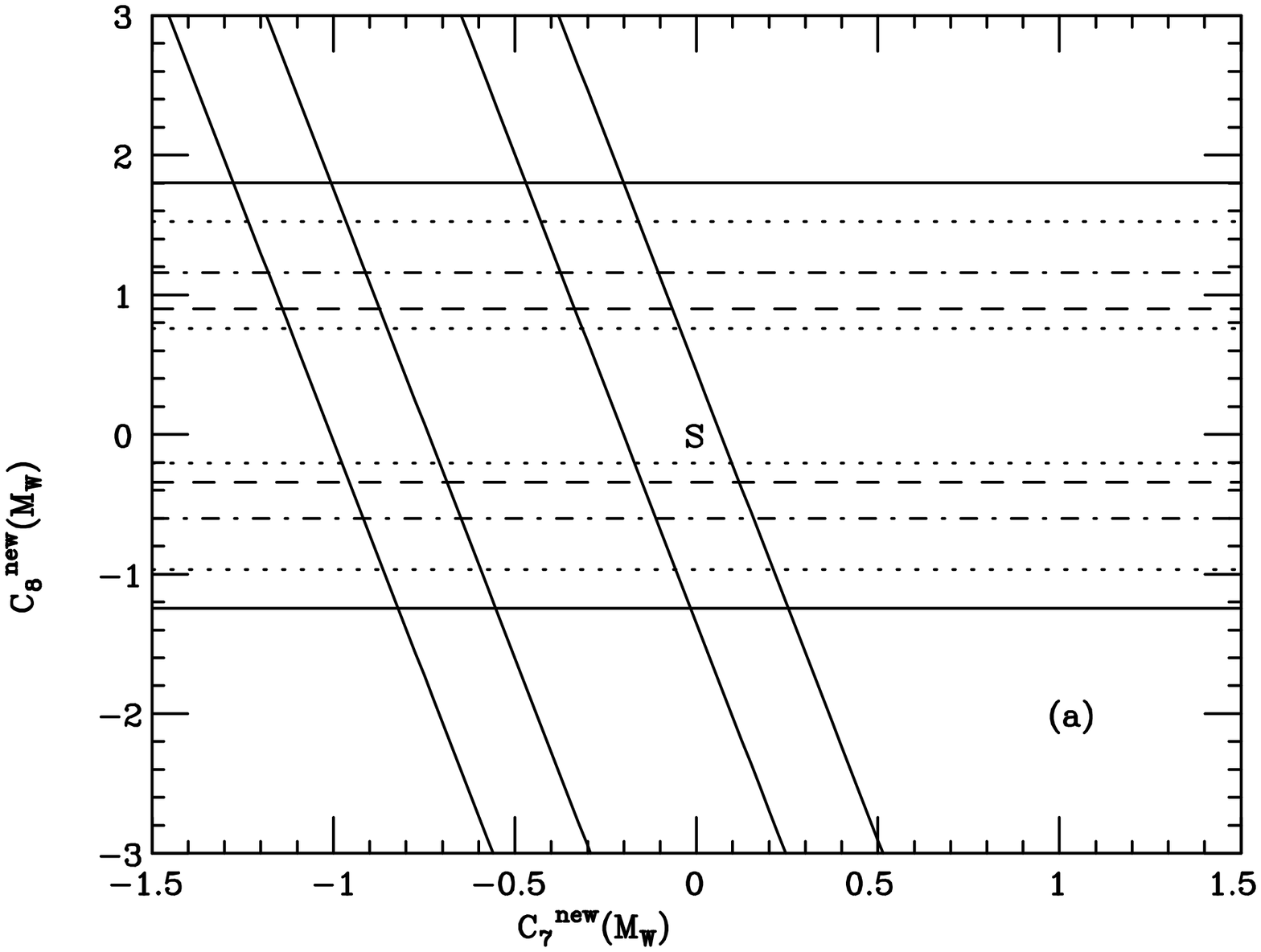,height=8.cm,width=8cm,angle=90}
\hspace*{-5mm}
\psfig{figure=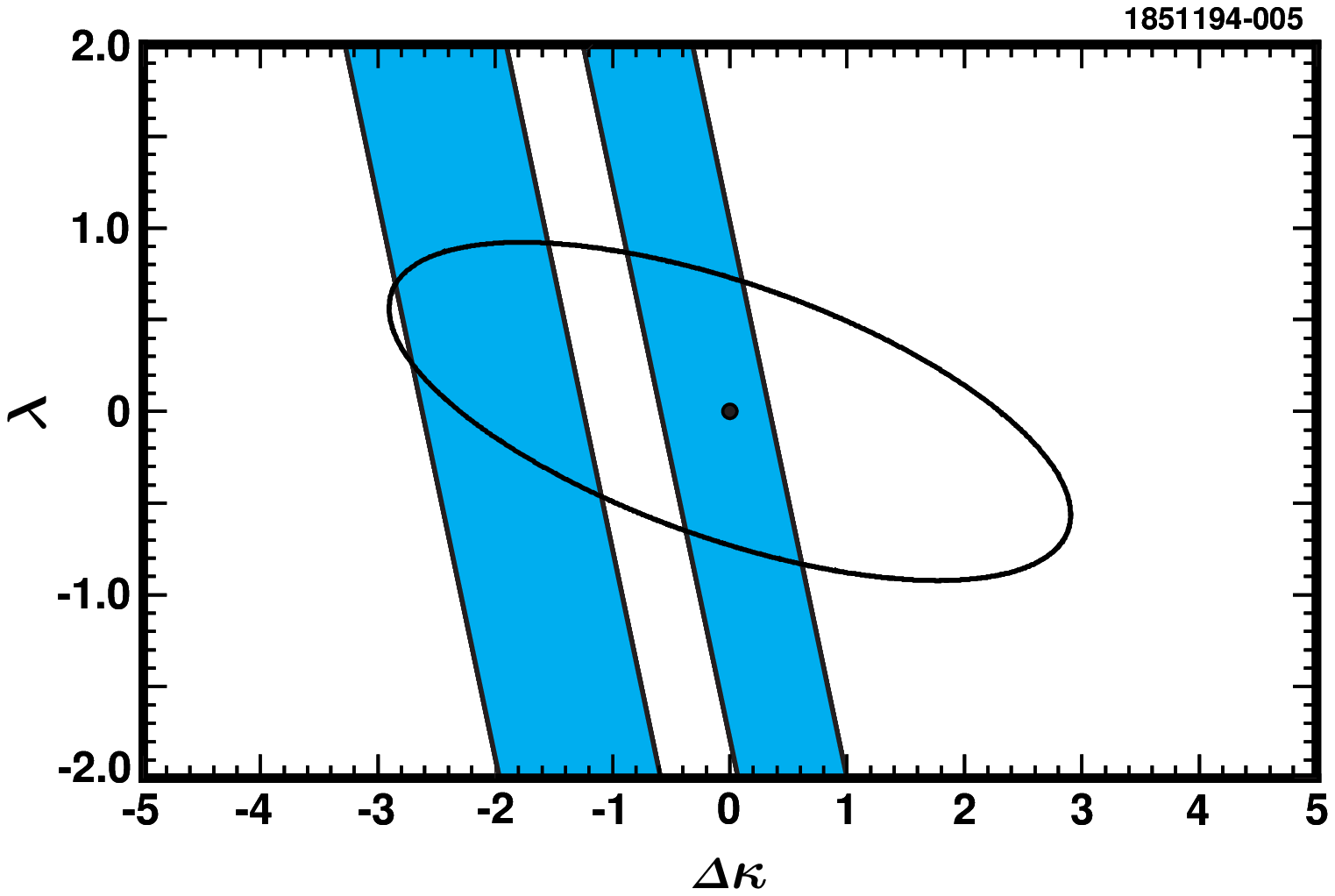,height=8.cm,width=8cm,angle=0}}
\vspace*{-0.75cm}
\centerline{
\psfig{figure=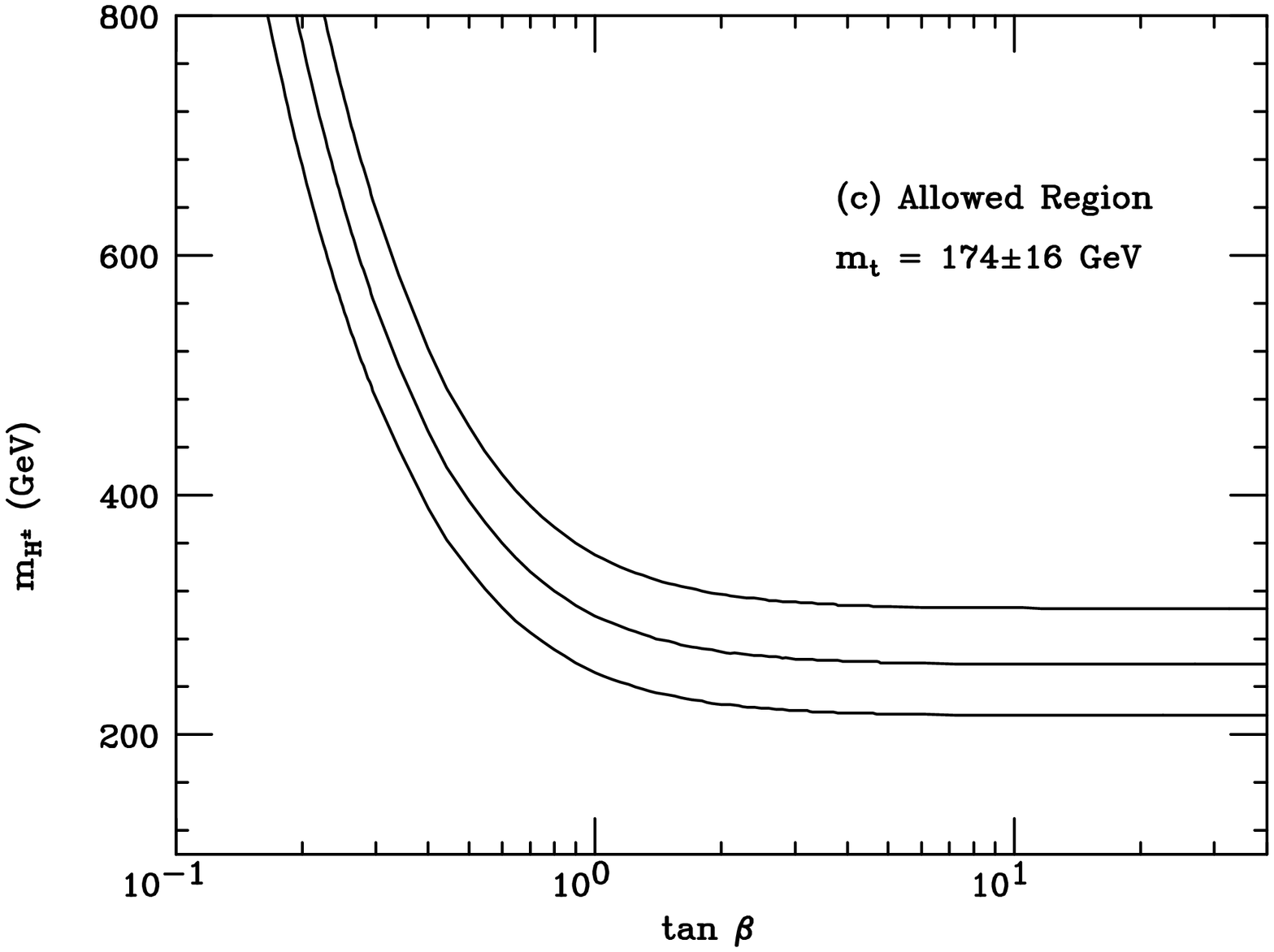,height=8.cm,width=8cm,angle=90}}
\vspace*{-1cm}
\caption{\small (a) Bounds on the contributions from new physics to $c_{7,8}$.
The region allowed by CLEO corresponds to the area inside the diagonal bands.
The horizontal lines represent potential measurements of $R\equiv
B(b\to sg)/B(b\to sg)_{SM}<30,20,10,5,3$ corresponding to the set of solid,
dotted, dash-dotted, dashed, and dotted lines, respectively.  The point
`S' represents the SM.  (b) Constraints on anomalous $WW\gamma$ couplings. The
shaded area is that allowed by CLEO and the interior of the ellipse is the
region allowed by D0.  The dot represent the SM values.
(c)  Limits from \bsg\
in the charged Higgs mass - \tb\ plane.  The excluded region is that to the
left and below the curves.  The three curves correspond to the values
$m_t=190, 174, 158\gev$ from top to bottom.}
\end{figure}

The inclusive process $b\to s\ell^+\ell^-$ also offers an excellent opportunity
to search for new physics.  The decay proceeds via electromagnetic and $Z$
penguin as well as by $W$ box diagrams, and hence can probe different coupling
structures than the pure electromagnetic process $b\to s\gamma$.
This reaction also receives long distance contributions
from the processes $B\rightarrow K^{(*)}\psi^{(')}$ followed by $\psi^{(')}
\rightarrow\ell^+\ell^-$ and from $c\bar c$ continuum intermediate states.
The short distance contributions  lead to the inclusive branching
fractions\cite{bsll} (including the leading logarithmic QCD corrections)
$B(B\rightarrow X_s\ell^+\ell^-)\sim (15, 7, 2)\times 10^{-6}$ for
$\ell=(e, \mu, \tau)$, respectively, and hence these modes will likely be
observed during the next few years.
The best method of separating the long and short
distance contributions, as well as observing any deviations from the
SM, is to measure the various kinematic distributions associated
with the final state lepton pair, such as the lepton pair invariant mass
distribution\cite{bsll}, the lepton pair forward-backward asymmetry\cite{ali},
and the tau polarization asymmetry\cite{mebsll} in the case $\ell=\tau$.
Measurement of all these quantities would allow for the determination of the
sign and magnitude of the Wilson coefficients for the electroweak loop
operators and thus provide a completely model independent analysis.
We note that measurement of these distributions requires the high
statistics samples which will be available at future B-factories.
The lepton pair invariant mass distribution for $b\to se^+e^-$ is displayed
in Fig. 3(a) (taking $m_t=175\gev$),
where the solid curve includes the contributions from the short
and long range effects and the dashed curve represents the short distance
alone.
We see that the long distance contributions
dominate only in the $M_{e^+e^-}$ regions near the $\psi$ and $\psi'$
resonances, and observations away from these peaks would cleanly separate the
short distance physics.  The tau polarization asymmetry is
presented in Fig. 3(b); we see that it is large and negative for this value
of $m_t$.  As an example of how new physics can affect this process, we
examine $b\to s\ell^+\ell^-$ in the case of an anomalous $WW\gamma$ vertex.
The resulting invariant mass spectrum is shown in Fig. 3(c) for several
values of $\Delta\kappa_\gamma$ (taking $\lambda_\gamma=0$), and the
variation of the tau polarization asymmetry with non-zero values of
$\Delta\kappa_\gamma$ and $\lambda_\gamma$ is displayed in Fig. 3(d)
for $\hat s\equiv q^2/m^2_b=0.7$.

\vspace*{-0.5cm}
\nn
\begin{figure}[htbp]
\centerline{
\psfig{figure=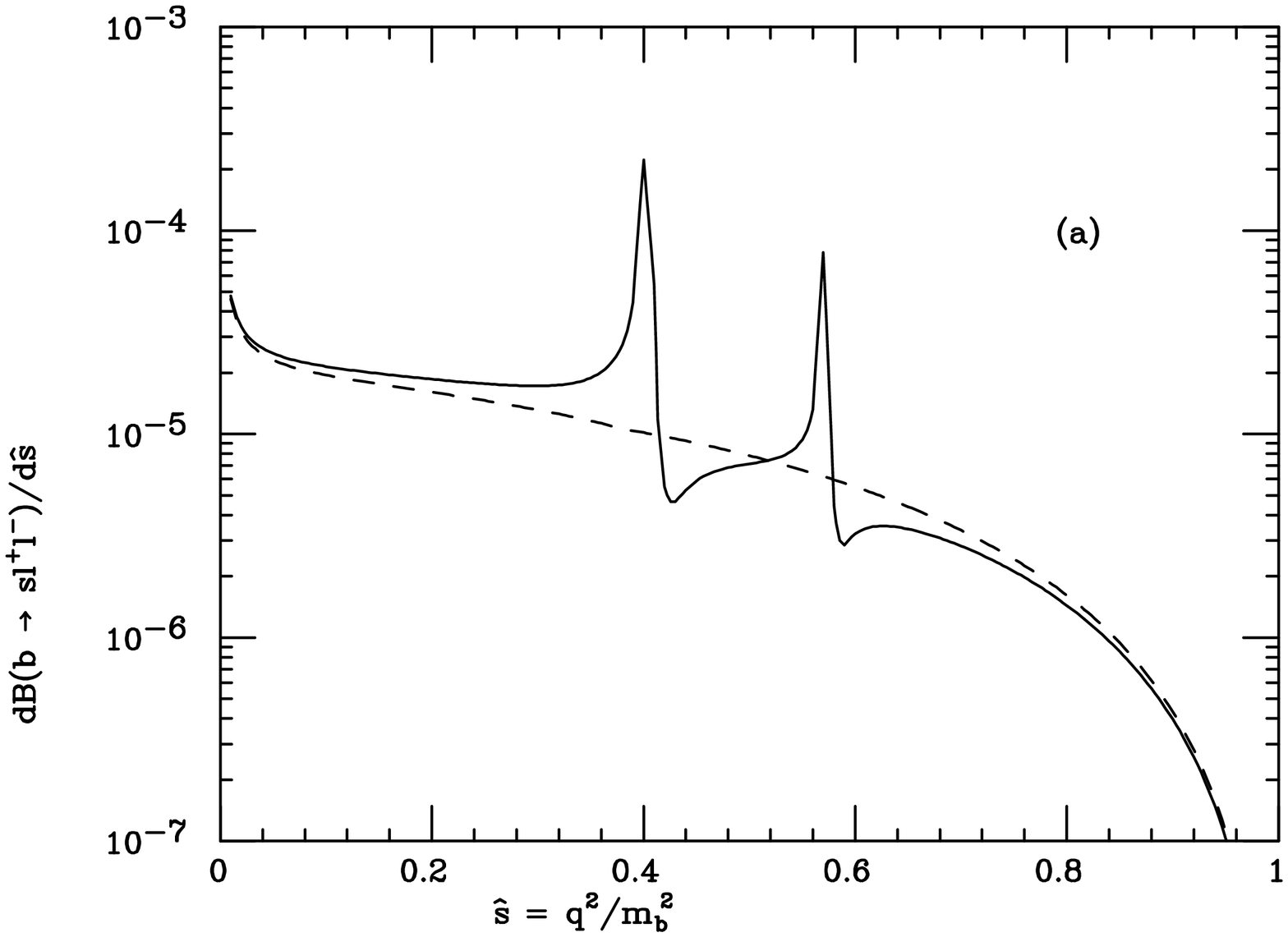,height=7.cm,width=8cm,angle=90}
\hspace*{-5mm}
\psfig{figure=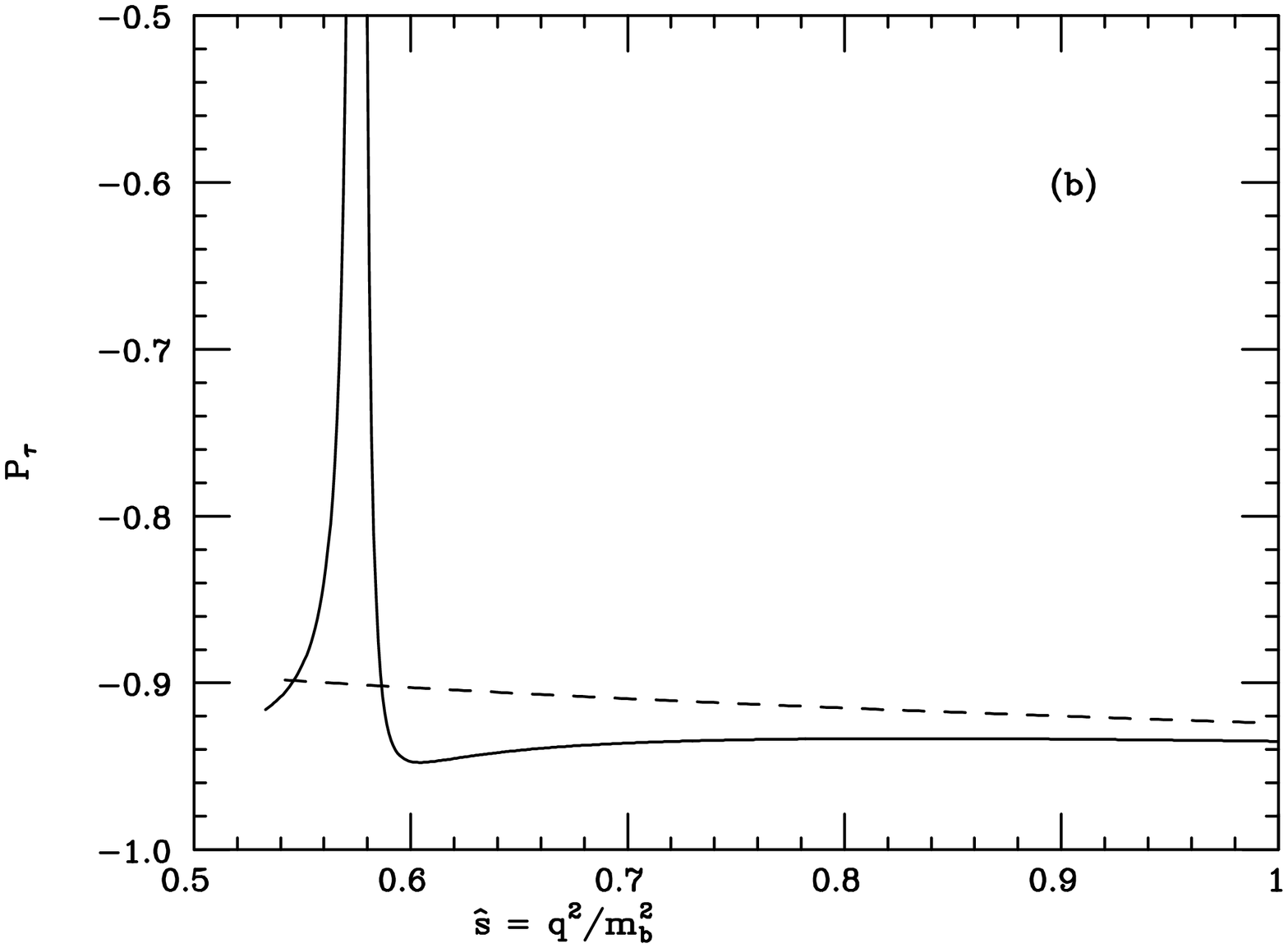,height=7.cm,width=8cm,angle=90}}
\vspace*{-0.75cm}
\centerline{
\psfig{figure=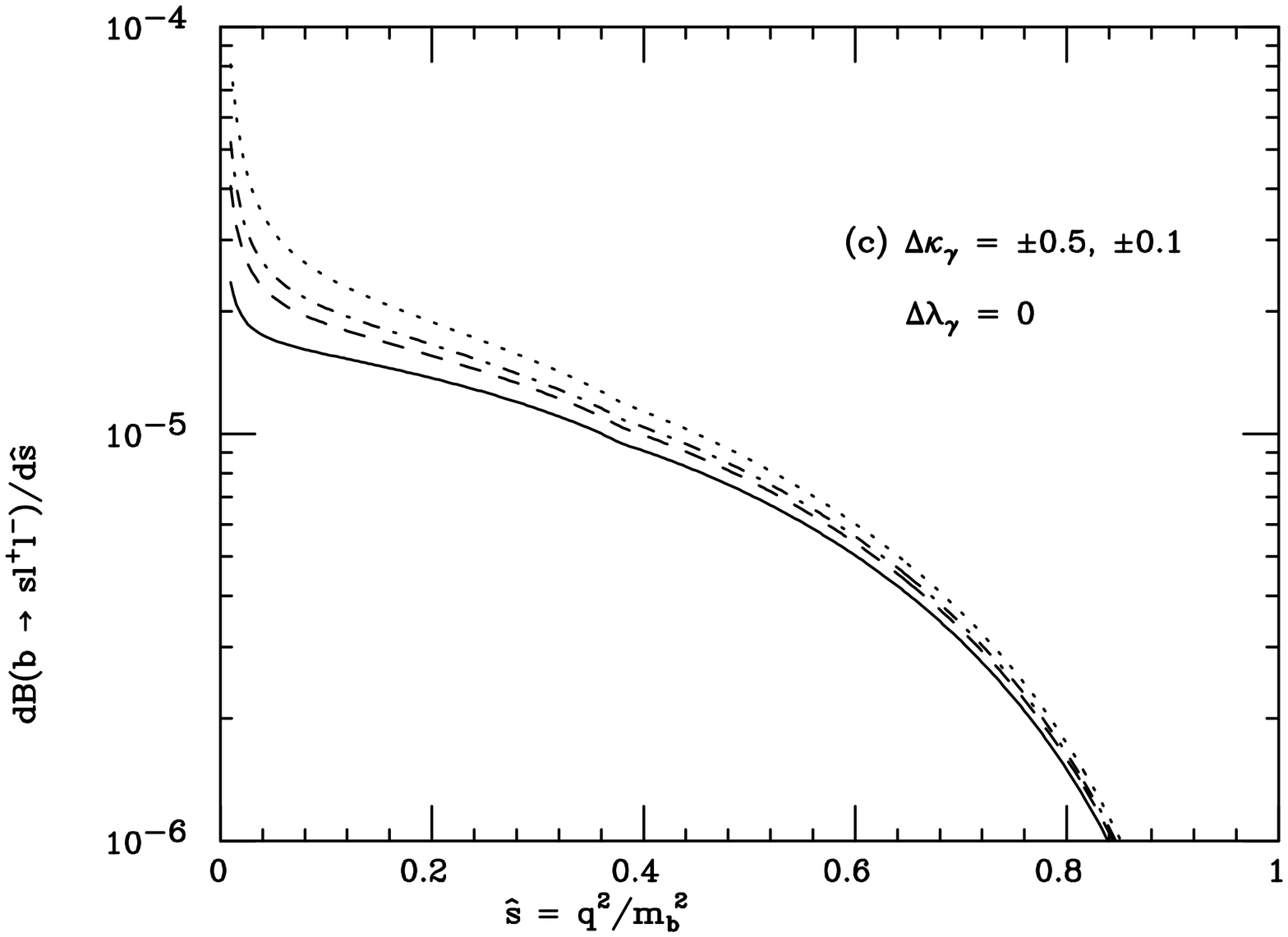,height=7.cm,width=8cm,angle=90}
\hspace*{-5mm}
\psfig{figure=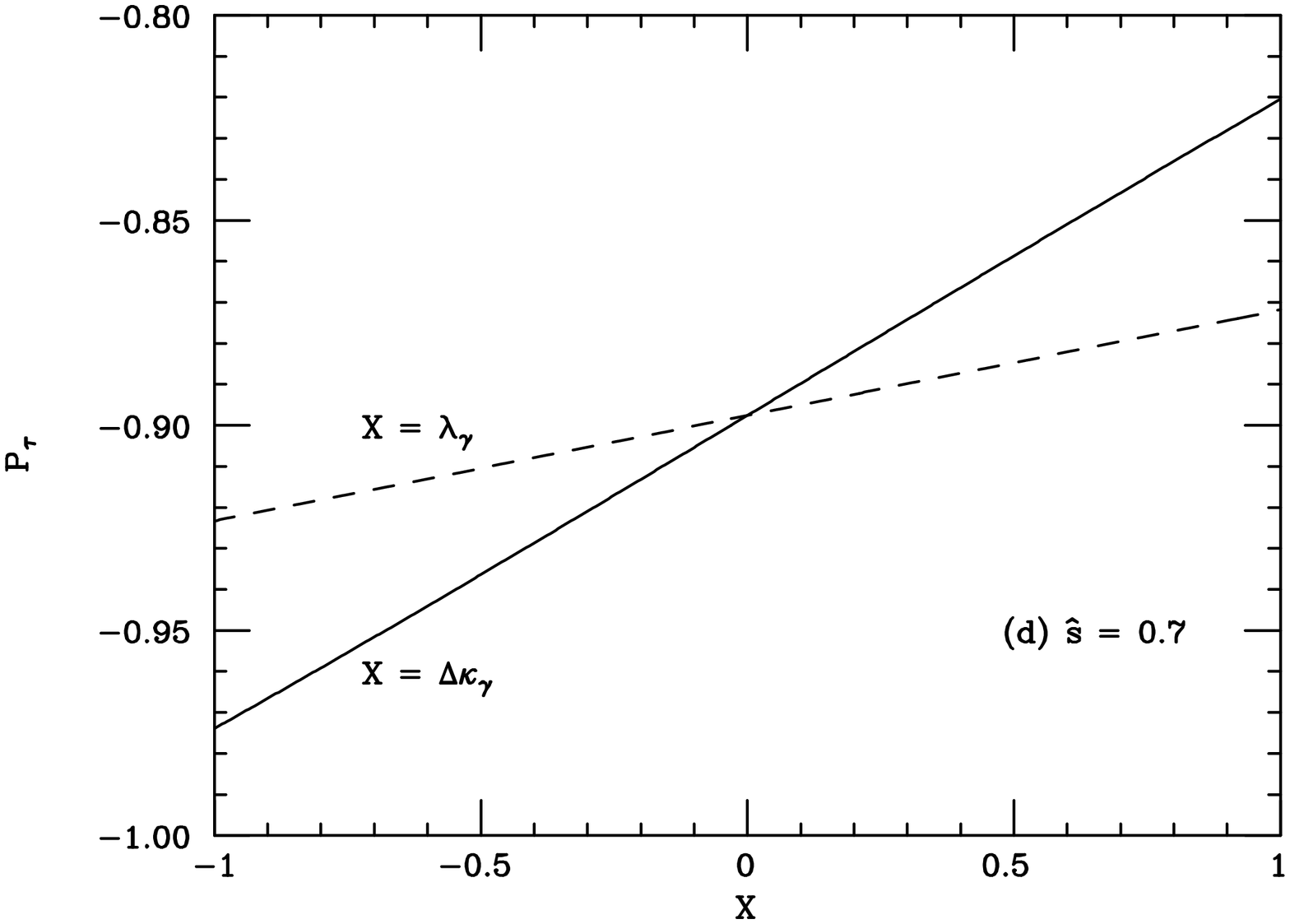,height=7.cm,width=8cm,angle=90}}
\vspace*{-1cm}
\caption{\small The (a) lepton pair mass distribution (with $\ell=e$) and (b)
tau polarization asymmetry (with $\ell=\tau$) in the SM, and the (c) lepton
pair mass distribution
and (d) tau polarization asymmetry with anomalous $WW\gamma$ couplings as
labeled, for the process $b\to s\ell^+\ell^-$ with $m_t=175\gev$.}
\end{figure}

In summary, we have seen that the process \bsg\ provides powerful constraints
for a variety of models containing physics beyond the SM.
In most cases, these constraints
either complement or are stronger than those from other
low-energy processes and from direct collider searches.  The decay $b\to
s\ell^+\ell^-$ is also an excellent probe of new physics.  It is sensitive
to possible new interactions since it allows the investigation
of various kinematic
distributions.  Measurement of these quantities would allow for
the determination of the sign and magnitude of all the contributing
Wilson coefficients.  We see that rare $B$ decays add powerful
insight to the quest for physics beyond the SM.
\vspace{1.0in}

{\elevenbf\noindent  References \hfil}
\vglue 0.2cm

\end{document}